\documentclass[runningheads]{llncs}
\usepackage[T1]{fontenc}

\usepackage[dvipsnames]{xcolor}
\usepackage{graphicx}
\usepackage{cite}
\usepackage{hyperref}

\usepackage{listings}%
\usepackage{indentfirst}%
\definecolor{codegreen}{rgb}{0,0.6,0}
\definecolor{codegray}{rgb}{0.5,0.5,0.5}
\definecolor{codepurple}{rgb}{0.58,0,0.82}
\definecolor{backcolour}{rgb}{0.95,0.95,0.92}%
\lstdefinestyle{mystyle}{
    backgroundcolor=\color{backcolour},   
    commentstyle=\color{codepurple},
    keywordstyle=\color{NavyBlue},
    numberstyle=\tiny\color{codegray},
    stringstyle=\color{codepurple},
    basicstyle=\ttfamily\footnotesize\bfseries,
    breakatwhitespace=false,         
    breaklines=true,                 
    captionpos=t,                    
    keepspaces=true,                 
    numbers=left,                    
    numbersep=5pt,                  
    showspaces=false,                
    showstringspaces=false,
    showtabs=false,                  
    tabsize=2
}%
\begin{document}
\title{\texttt{MechIR}: A Mechanistic Interpretability Framework for Information Retrieval}
\titlerunning{MechIR}
\author{Andrew Parry\inst{1}\orcidID{0000-0001-5446-8328} \and
Catherine Chen\inst{2}\orcidID{0009-0009-8734-436X} \and
Carsten Eickhoff \inst{3}\orcidID{0000-0001-9895-4061} \and Sean MacAvaney\inst{1}\orcidID{0000-0002-8914-2659}}
\authorrunning{Parry et al.}
\institute{University of Glasgow, Glasgow, Scotland, UK \and
Brown University, Providence, RI, USA \and
University of Tübingen, Tübingen, Germany}
\maketitle              %
\begin{abstract}
Mechanistic interpretability is an emerging diagnostic approach for neural models that has gained traction in broader natural language processing domains. This paradigm aims to provide attribution to components of neural systems where causal relationships between hidden layers and output were previously uninterpretable. As the use of neural models in IR for retrieval and evaluation becomes ubiquitous, we need to ensure that we can interpret \textit{why} a model produces a given output for both transparency and the betterment of systems. This work comprises a flexible framework for diagnostic analysis and intervention within these highly parametric neural systems specifically tailored for IR tasks and architectures. In providing such a framework, we look to facilitate further research in interpretable IR with a broader scope for practical interventions derived from mechanistic interpretability. We provide preliminary analysis and look to demonstrate our framework through an axiomatic lens to show its applications and ease of use for those IR practitioners inexperienced in this emerging paradigm.


\vspace{0.6em}
\hspace{5.5em}\includegraphics[width=1.25em,height=1.25em]{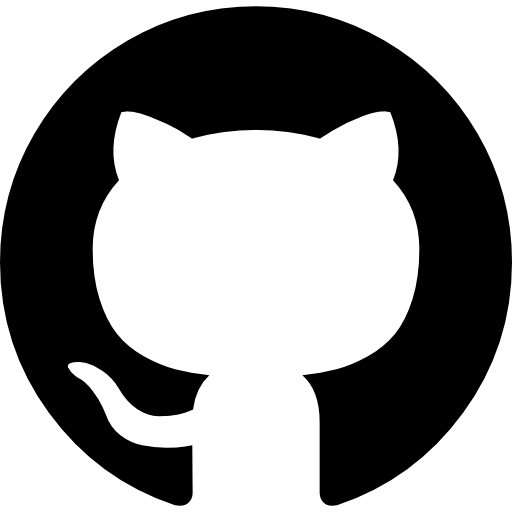}\hspace{.3em}
\parbox[c]{\columnwidth}
{
    \vspace{-.55em}
    \href{https://github.com/Parry-Parry/MechIR}{\nolinkurl{https://github.com/Parry-Parry/MechIR}}
}
\vspace{-1.2em}

\end{abstract}
\section{Introduction}

In recent years, machine learning models have achieved high performance, leading to widespread adoption and integration into tools used across various fields and industries. However, the internal decision-making processes of these models remain opaque due to their deep parametric nature, which makes translating model behavior into human-understandable terms difficult. This can be particularly concerning in safety-sensitive domains, such as healthcare and law, where understanding model behavior is critical for detecting and correcting potential errors. Without effective tools to investigate these complex models, our ability to intuitively understand and explain their behavior is limited.

However, recent mechanistic interpretability methods offer a promising solution for understanding the internal mechanisms of neural networks by performing causal interventions on specific model components \cite{vig:2020, bereska2024mechanistic}. Using causal methods such as activation patching, researchers have identified components responsible for generative language modelling tasks such as indirect object identification \cite{wang2022interpretability} and applied learned insights to improve model performance \cite{meng2022locating, merullo:2024}. Much of this progress in the broader NLP community is driven by the availability of open-source tooling and tutorials, such as TransformerLens \cite{nanda2022transformerlens}. However, current resources are tailored for generative language tasks. To make mechanistic interpretability research accessible in IR, we present \texttt{MechIR}, a Python package that aims to help researchers reverse-engineer IR models and causally investigate internal model components. 

\begin{figure}[tb]
    \centering
    \includegraphics[width=\linewidth]{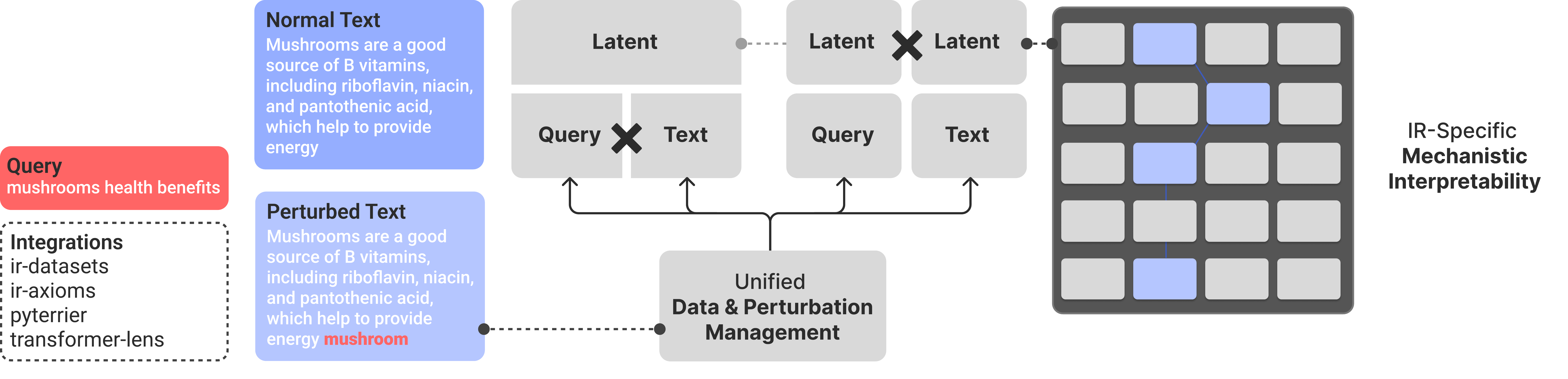}
    \caption{MechIR allows for common IR architectures to be analyzed under text perturbation. Here a query terms is added to the text and we can observe the attention heads which respond to the addition of this term. Both cross-encoders and bi-encoders can be analyzed.}
    \label{fig:pipeline-overview}
\end{figure}

\section{Background}

\subsubsection{Activation Patching}
Activation patching is a causal intervention method that aims to localize the specific component(s) responsible for a targeted behavior \cite{geiger2021causal, vig:2020, meng2022locating}. In IR, this approach involves a pair of inputs, consisting of one perturbed input($X_{perturbed}$) and one baseline input ($X_{baseline}$). The perturbed input is constructed by applying some function to modify an original input (e.g., inserting a query term to the end of a document) and the baseline input is a padded variant of the original input to maintain token lengths between the pairs \cite{chen2024axiomatic}. These pairs are then used in three forward passes: 
\begin{enumerate}
    \item \textit{Baseline run:} run the model on $X_{baseline}$, recording model performance and caching activations (if expected performance is greater than $X_{perturbed}$).
    \item \textit{Perturbed run:} run the model on $X_{perturbed}$, recording model performance and caching activations (if expected performance is greater than $X_{baseline}$).
    \item \textit{Patched run:} run the model on $X_{baseline}$ or $X_{perturbed}$ (whichever has the lower expected performance) with model component(s) replaced from the cached activations from the other run and record the final model performance.
\end{enumerate}
Model performance is defined by the relevance score for an input, with a specific measure depending on the model type. For example, the similarity score (e.g., dot product) between query and document representations is often used as the relevance score in bi-encoders, whereas logits are typically used in cross-encoders. To evaluate the effect of a patch and determine which component(s) are responsible for the difference in model behavior, we examine changes to the model's performance during the \textit{patched run}. For more details, we refer the reader to Chen et al.~\cite{chen2024axiomatic} and the supplementary demo notebooks.

\subsubsection{TransformerLens}
TransformerLens \cite{nanda2022transformerlens} is an open-source Python package for analyzing decoder-only Transformer-based language models, widely used in mechanistic interpretability research for generative language models \cite{nanda2023progress, conmy2023towards, gurnee2023finding}. Researchers can cache, edit, remove, or replace model activations to explore how specific model components influence overall behavior. \texttt{MechIR} extends this functionality to neural IR models and additionally offers integration with widely used IR research tools such as IR datasets \cite{macavaney:sigir2021-irds} and PyTerrier \cite{pyterrier2020ictir}. Overall, our package is an end-to-end tool for mechanistic interpretability research in IR.

\section{\texttt{MechIR} Overview}

In this section, we provide an overview of the \texttt{MechIR} framework (Figure \ref{fig:pipeline-overview})  and supported functionality.

\subsubsection{Accessing Model Activations}
TransformerLens enables access to a transformer's internal activations by re-implementing the model with hooks on components of interest (i.e., attention heads and MLPs). In \texttt{MechIR}, we extend support to common retrieval architectures, including bi-encoders (e.g., TAS-B \cite{hofstatter2021efficiently}) and cross-encoders (e.g., monoELECTRA \cite{pradeep2022squeezing}). A full list of supported architectures can be found in the package documentation. Below observe the instantiation of a bi-encoder in MechIR, the underlying interface infers the original model architecture and converts accordingly.

\begin{lstlisting}[language=Python]
from mechir import Dot
model = Dot(``bert-base-uncased'')
all_hooks = model.hook_dict
\end{lstlisting}

\subsubsection{Activation Patching}
\texttt{MechIR} supports activation patching for the following components: (1) \textit{blocks:} patches components in an entire layer (e.g., residual stream, attention outputs, MLP layer) across individual token positions in the input pairs and (2) \textit{attention heads:} patches attention heads in specific layers across all or individual token positions. Additionally, we provide basic plotting functionality to support analysis on experimental results, specifically to visualize the results of activation patching and attention patterns.

\subsubsection{Paired Dataset Creation}
Activation patching relies on pairs of inputs constructed by perturbing an original query-document pair. To create this dataset, we provide support for loading from IR datasets \cite{macavaney:sigir2021-irds} and a \textit{perturbation} class that allows users to create their input pairs. Provided supplementary tutorial materials discuss possible general perturbation methods and outline an analysis pipeline using our framework. Below we define a perturbation which can be added to a torch dataloader through collate functions.

\begin{lstlisting}[language=Python]
  from mechir.data import MechIRDataset, DotDataCollator
  from mechir.perturb import perturbation
  from torch.utils.data import DataLoader
  
  @perturbation
  def func(document):
     return ``MechIR! '' + document
  dataset = MechIRDataset(``msmarco-passage'')
  collator = DotDataCollator(model.tokenizer, func)
  loader = DataLoader(dataset, collate_fn=collator)
  outputs = [model.patch(**b) for b in loader]
\end{lstlisting}

\subsubsection{Open Source Materials}
We provide our full resource in an open-source Python package, which can be installed via \texttt{pip install mechir}. Additionally, we provide documentation and several supplementary tutorial notebooks in the source code repository. These tutorials cater to both novice and experienced researchers in interpretability, covering (1) an example of an end-to-end experimental pipeline (demonstration) and (2) considerations for activation patching in IR. We plan to continue development and add additional functionality for other mechanistic interpretability methods (e.g., path patching) in the future. 

\section{Demonstration}

\begin{figure}[tb]
    \centering
    \includegraphics[width=\textwidth]{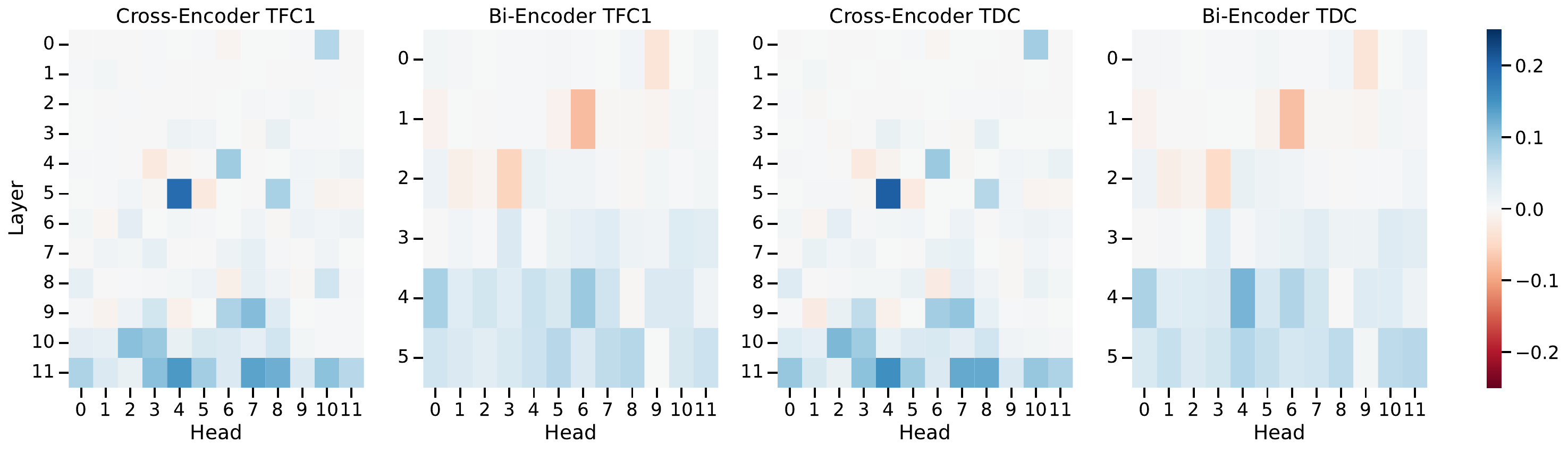}
    \caption{Effect of inserting different types of query terms to documents (\textit{left}: TFC1, \textit{right}: TDC) with activation patching over a bi- and cross-encoder for a subsample of texts with a ``highly-relevant'' judgment.}
    \label{fig:axiom}
\end{figure}

Our demonstration presents a concrete application of our framework, showing all aspects of an investigation of relevance approximation through activation patching. We provide an introduction to mechanistic interpretability using common IR architectures, aiming to provide intuitions behind this research area while grounding this introduction in familiar concepts (ranking axioms).

We will guide users through the process of loading a dataset into MechIR before choosing a perturbation to create input pairs that aim to isolate behavior to particular components of neural ranking models. We take a relevance grade-stratified subsample ($n=1000$) of the TREC DL19 and DL20 test collection and observe how different IR model architectures (cross-encoder vs. bi-encoder) behave under the addition of query terms. We perform head patching to see how the addition of random query terms (TFC1 \cite{fang:2004}) versus the most discriminative terms by IDF (TDC \cite{fang:2011}) changes model behaviour.

Observe in Figure \ref{fig:axiom} the diffuse nature of bi-encoder activations under term addition; no one head strongly activates under term matching with different heads in later layers activating for any query term versus discriminative query terms. This is somewhat intuitive as there is no term-level interactions between queries and documents within this archicture. Additionally, the addition of these tokens frequently is penalized in earlier layers; whether or not this is an artifact of term matching or positional bias would require future investigation\footnote{See Jiang et al.~\cite{jiang:2021} and Parry et al.~\cite{parry:2024} for treatments of these phenomena}. In contrast, within a cross-encoder, several heads activate consistently under term matching with slight increases in activations dependent on the salience of terms; the majority of strong activations are observed in the final layers aligning with prior post-hoc probing by Jiang et al. \cite{jiang:2021} but with some strong activations in the middle layers particularly under TDC. The key difference in validating prior findings with these methods is that by isolating this behavior to particular components, we can intervene, for example, to reduce bi-encoder penalization of salient terms.

\section{Target Audience and Potential Use Cases}

The target audience for this demo is PhD students and researchers working on explainable information retrieval (XIR) or those interested in starting interpretability research. While the functionality of this first iteration of \texttt{MechIR} is focused on neural search, we hope it sparks collaborative efforts to extend mechanistic interpretability to other areas of IR, such as recommender systems. Overall, this line of research offers exciting opportunities, including the development of diagnostic tools to enhance model performance, mitigate bias, prevent adversarial attacks, and create steerable and personalized systems.

\begin{credits}
\subsubsection{\ackname} We thank Jack Merullo and Gregory Polyakov for their comments and suggestions on this work.

\subsubsection{\discintname}
The authors have no competing interests to declare that are relevant to the content of this article.
\end{credits}

\bibliographystyle{splncs04}%
\bibliography{refs}

\begin{thebibliography}{10}
\providecommand{\url}[1]{\texttt{#1}}
\providecommand{\urlprefix}{URL }
\providecommand{\doi}[1]{https://doi.org/#1}

\bibitem{bereska2024mechanistic}
Bereska, L., Gavves, E.: Mechanistic interpretability for ai safety--a review. arXiv preprint arXiv:2404.14082  (2024)

\bibitem{chen2024axiomatic}
Chen, C., Merullo, J., Eickhoff, C.: Axiomatic causal interventions for reverse engineering relevance computation in neural retrieval models. In: Proceedings of the 47th International ACM SIGIR Conference on Research and Development in Information Retrieval. pp. 1401--1410 (2024)

\bibitem{conmy2023towards}
Conmy, A., Mavor-Parker, A., Lynch, A., Heimersheim, S., Garriga-Alonso, A.: Towards automated circuit discovery for mechanistic interpretability. Advances in Neural Information Processing Systems  \textbf{36},  16318--16352 (2023)

\bibitem{fang:2004}
Fang, H., Tao, T., Zhai, C.: A formal study of information retrieval heuristics. In: Sanderson, M., J{\"{a}}rvelin, K., Allan, J., Bruza, P. (eds.) {SIGIR} 2004: Proceedings of the 27th Annual International {ACM} {SIGIR} Conference on Research and Development in Information Retrieval, Sheffield, UK, July 25-29, 2004. pp. 49--56. {ACM} (2004). \doi{10.1145/1008992.1009004}, \url{https://doi.org/10.1145/1008992.1009004}

\bibitem{fang:2011}
Fang, H., Tao, T., Zhai, C.: Diagnostic evaluation of information retrieval models. {ACM} Trans. Inf. Syst.  \textbf{29}(2),  7:1--7:42 (2011). \doi{10.1145/1961209.1961210}

\bibitem{geiger2021causal}
Geiger, A., Lu, H., Icard, T., Potts, C.: Causal abstractions of neural networks. Advances in Neural Information Processing Systems  \textbf{34},  9574--9586 (2021)

\bibitem{gurnee2023finding}
Gurnee, W., Nanda, N., Pauly, M., Harvey, K., Troitskii, D., Bertsimas, D.: Finding neurons in a haystack: Case studies with sparse probing. arXiv preprint arXiv:2305.01610  (2023)

\bibitem{hofstatter2021efficiently}
Hofst{\"a}tter, S., Lin, S.C., Yang, J.H., Lin, J., Hanbury, A.: Efficiently teaching an effective dense retriever with balanced topic aware sampling. In: Proceedings of the 44th International ACM SIGIR Conference on Research and Development in Information Retrieval. pp. 113--122 (2021)

\bibitem{jiang:2021}
Jiang, Z., Tang, R., Xin, J., Lin, J.: How does {BERT} rerank passages? an attribution analysis with information bottlenecks. In: Bastings, J., Belinkov, Y., Dupoux, E., Giulianelli, M., Hupkes, D., Pinter, Y., Sajjad, H. (eds.) Proceedings of the Fourth BlackboxNLP Workshop on Analyzing and Interpreting Neural Networks for NLP. pp. 496--509. Association for Computational Linguistics, Punta Cana, Dominican Republic (Nov 2021). \doi{10.18653/v1/2021.blackboxnlp-1.39}, \url{https://aclanthology.org/2021.blackboxnlp-1.39}

\bibitem{macavaney:sigir2021-irds}
MacAvaney, S., Yates, A., Feldman, S., Downey, D., Cohan, A., Goharian, N.: Simplified data wrangling with ir\_datasets. In: SIGIR (2021)

\bibitem{pyterrier2020ictir}
Macdonald, C., Tonellotto, N.: Declarative experimentation information retrieval using pyterrier. In: Proceedings of ICTIR 2020 (2020)

\bibitem{meng2022locating}
Meng, K., Bau, D., Andonian, A., Belinkov, Y.: Locating and editing factual associations in gpt. Advances in Neural Information Processing Systems  \textbf{35},  17359--17372 (2022)

\bibitem{merullo:2024}
Merullo, J., Eickhoff, C., Pavlick, E.: Circuit component reuse across tasks in transformer language models. In: The Twelfth International Conference on Learning Representations (2024), \url{https://openreview.net/forum?id=fpoAYV6Wsk}

\bibitem{nanda2022transformerlens}
Nanda, N., Bloom, J.: Transformerlens. \url{https://github.com/neelnanda-io/TransformerLens} (2022)

\bibitem{nanda2023progress}
Nanda, N., Chan, L., Lieberum, T., Smith, J., Steinhardt, J.: Progress measures for grokking via mechanistic interpretability. arXiv preprint arXiv:2301.05217  (2023)

\bibitem{parry:2024}
Parry, A., MacAvaney, S., Ganguly, D.: Exploiting positional bias for query-agnostic generative content in search. In: Ku, L.W., Martins, A., Srikumar, V. (eds.) Findings of the Association for Computational Linguistics ACL 2024. pp. 11030--11047. Association for Computational Linguistics, Bangkok, Thailand and virtual meeting (Aug 2024). \doi{10.18653/v1/2024.findings-acl.656}, \url{https://aclanthology.org/2024.findings-acl.656}

\bibitem{pradeep2022squeezing}
Pradeep, R., Liu, Y., Zhang, X., Li, Y., Yates, A., Lin, J.: Squeezing water from a stone: a bag of tricks for further improving cross-encoder effectiveness for reranking. In: European Conference on Information Retrieval. pp. 655--670. Springer (2022)

\bibitem{vig:2020}
Vig, J., Gehrmann, S., Belinkov, Y., Qian, S., Nevo, D., Singer, Y., Shieber, S.: Investigating gender bias in language models using causal mediation analysis. Advances in neural information processing systems  \textbf{33},  12388--12401 (2020)

\bibitem{wang2022interpretability}
Wang, K., Variengien, A., Conmy, A., Shlegeris, B., Steinhardt, J.: Interpretability in the wild: a circuit for indirect object identification in gpt-2 small. arXiv preprint arXiv:2211.00593  (2022)

\end{thebibliography}

\end{document}